\documentclass[12pt,draftcls,onecolumn]{IEEEtran}

\IEEEoverridecommandlockouts

\usepackage{amssymb,amsmath}
\usepackage{graphicx}
\usepackage{tikz}
\usepackage{color}

\definecolor{verm}{rgb}{0.6,0.2,0.2}
\definecolor{purp}{rgb}{0.3,0.1,0.6}
\definecolor{purple}{rgb}{0.4,0.0,0.6}
\definecolor{bggreen}{rgb}{0.1,0.3,0.1}
\definecolor{dgreen}{rgb}{0.1,0.6,0.1}
\definecolor{black}{rgb}{0.0,0.0,0.0}
\definecolor{crim}{rgb}{0.3,0.1,0.1}
\definecolor{dred}{rgb}{0.5,0.1,0.1}

\def\a{{\bf a}}
\def\b{{\bf b}}

\def\eb{{\bf e}}

\def\h{{\bf h}}

\def\v{{\bf v}}

\def\x{{\bf x}}
\def\y{{\bf y}}
\def\z{{\bf z}}

\def\B{{\cal B}}

\def\F{{\cal F}}

\def\M{{\cal M}}
\def\N{{\cal N}}

\def\Abb{{\mathbb A}}
\def\Bbb{{\mathbb B}}
\def\Cbb{{\mathbb C}}

\def\Nbb{{\mathbb N}}

\def\R{{\mathbb R}}
\def\Sm{{\mathbb S}}

\def\al{\alpha}
\def\d{\delta}
\def\D{\Delta}
\def\e{\epsilon}
\def\g{\gamma}
\def\GA{\Gamma}
\def\l{\lambda}

\def\OM{\Omega}
\def\r{\rho}
\def\s{\sigma}
\def\SI{\Sigma}
\def\t{\tau}
\def\th{\theta}
\def\Th{\Theta}

\def\boldeta{{\boldsymbol \eta}}
\def\bdelta{{\boldsymbol \d}}

\def\bmu{{\boldsymbol \mu}}
\def\bnu{{\boldsymbol \nu}}
\def\bth{{\boldsymbol \theta}}
\def\bphi{{\boldsymbol \phi}}
\def\bpsi{{\boldsymbol \psi}}

\def\bxi{{\boldsymbol \xi}}
\def\bzeta{{\boldsymbol \zeta}}

\def\lai{l \ap \infty}
\def\mai{m \ap \infty}
\def\nai{n \ap \infty}

\def\ap{\rightarrow}

\def\seq{\subseteq}

\def\bi{\{0,1\}}

\def\bp{\{-1,1\}}
\def\bz{{\bf 0}}
\def\imp{\; \Rightarrow \;}

\def\fa{\; \forall}

\def\st{\mbox{ s.t. }}

\def\nm{\Vert}

\renewcommand{\iff}{\mbox{$\; \; \Leftrightarrow \; \;$}}
\renewcommand{\and}{\mbox{$\wedge$}}

\newcommand{\bc}{\begin{center}}
\newcommand{\ec}{\end{center}}
\newcommand{\be}{\begin{equation}}
\newcommand{\ee}{\end{equation}}
\newcommand{\bd}{\begin{displaymath}}
\newcommand{\ed}{\end{displaymath}}
\newcommand{\ba}{\begin{array}}
\newcommand{\ea}{\end{array}}
\newcommand{\ben}{\begin{enumerate}}
\newcommand{\een}{\end{enumerate}}
\newcommand{\bit}{\begin{itemize}}
\newcommand{\eit}{\end{itemize}}
\newcommand{\beq}{\begin{eqnarray}}
\newcommand{\eeq}{\end{eqnarray}}
\newcommand{\btab}{\begin{tabular}}
\newcommand{\etab}{\end{tabular}}
\newcommand{\bfig}{\begin{figure}}
\newcommand{\efig}{\end{figure}}

\def\alh{\hat{\al}}
\def\betah{\hat{\beta}}
\def\phih{\hat{\phi}}
\def\Phih{\hat{\Phi}}

\def\halmos{\hfill\mbox{$\blacksquare$}}

\newtheorem{corollary}{Corollary}{\bf}{\it}
\newtheorem{definition}{Definition}{\bf}{\it}
{\bf}{\rm}
{\bf}{\it}
\newtheorem{theorem}{Theorem}{\bf}{\it}
{\bf}{\it}
{\bf}{\it}
{\bf}{\rm}

\begin{document}

\title{
Mixing Coefficients Between \\
Discrete and Real Random Variables:\\
Computation and Properties
}
\author{
Mehmet Eren Ahsen and M.\ Vidyasagar
\thanks{Department of Bioengineering,
University of Texas at Dallas, 800 W.\ Campbell Road, Richardson, TX 75080;
Emails: \{Ahsen, M.Vidyasagar\}@utdallas.edu.
This work is supported by National Science Foundation
Award \#1001643.}
}
\maketitle

\begin{abstract}

In this paper we study the problem of estimating the 
alpha-, beta- and phi-mixing coefficients
between two random variables, that 
can either assume values in a finite set or the set of
real numbers.
In either case, explicit closed-form formulas for the beta-mixing
coefficient are already known.
Therefore for random variables assuming values in a finite set,
our contributions are two-fold: (i) In the case of the
alpha-mixing coefficient, we show that determining whether or not
it exceeds a prespecified threshold is NP-complete,
and provide efficiently computable upper and lower bounds.
(ii) We derive an exact closed-form formula for the phi-mixing coefficient.
Next, we prove analogs of the data-processing inequality from information
theory for each of the three kinds of mixing coefficients.
Then we move on to real-valued random variables, and show that 
by using percentile binning and allowing the number of bins
to increase more slowly than the number of samples, we can generate
empirical estimates that are consistent, i.e., converge to the true values
as the number of samples approaches infinity.

\end{abstract}

\section{Introduction}\label{sec:1}

The notion of independence of random variables is central to probability
theory.
In \cite[p.\ 8]{Kolmogorov33}, Kolmogorov says:
\begin{quote}
``Indeed, as we have already seen, the theory of probability can be
regarded from the mathematical point of view as a special application
of the general theory of additive set functions.
\end{quote}
and
\begin{quote}
``Historically, the independence of experiments and random variables represents
the very mathematical concept that has given the theory of probability
its peculiar stamp.''
\end{quote}
In effect, Kolmogorov is saying that, if the notion of independence
is removed, then probability theory reduces to just measure theory.

Independence is a binary concept: 
Either two random variables are independent, or they are not.
It is therefore worthwhile to replace the concept of independence
with a more nuanced measure that {\it quantifies\/} the extent to which
given random variables are dependent.
In the case of stationary stochastic processes, there are various
notions of `mixing', corresponding to long term asymptotic independence.
These notions can be readily adapted to define various mixing coefficients
between two random variables.
In this setting, the mixing rate of a stochastic process can
be interpreted as the mixing coefficient between the semi-infinite
`past' and `future' variables.
Several such definitions are presented in \cite[p.\ 3]{Doukhan94}, 
out of which three are of interest to us, namely the $\al$-, $\beta$-
and $\phi$-mixing coefficients.
While the definitions themselves are well-known, there is very little
work on actually {\it computing\/} (or at least estimating) these mixing
coefficients in a given situation.
The $\beta$-mixing coefficient is easy to compute but this is not
the case for the $\al$- and the $\phi$-mixing coefficients.

Against this background, the present paper makes the following
specific contributions:
For random variables that assume values in a finite set:
\ben
\item
In the case of the $\al$-mixing coefficient,
it is shown that determining whether or not it exceeds a prespecified
threshold is NP-complete, and
efficiently computable upper and lower bounds are derived.
\item An efficiently computable
exact formula is derived for the $\phi$-mixing coefficient.
\item We study the case of three random variables $X,Y,Z$, where
$X,Z$ are conditionally independent given $Y$,
or equivalently, $X \ap Y \ap Z$ is a short Markov chain.
In this case a well-known inequality from information theory
\cite[p.\ 34]{CT06}, usually referred to as the `data processing inequality
(DPI)', states that
\be\label{eq:11}
I(X,Z) \leq \min \{ I(X,Y) , I(Y,Z) \} ,
\ee
where $I(\cdot,\cdot)$ denotes the mutual information.
We state and prove analogs of the DPI for each of the $\al$-, $\beta$-
and $\phi$-mixing coefficients.
\een
Next we turn to real-valued random variables.
\ben
\item Suppose $X,Y$ are real-valued random variables whose joint
distribution has a density with respect to the Lebesgue measure, and that
$\{ (x_1 , y_1 ) , \ldots, ( x_l , y_l ) \}$ are independent
samples of $(X,Y)$.
If we compute the empirical joint distribution of $(X,Y)$ from these
samples, then the Glivenko-Cantelli Lemma states that the empirical
joint distribution converges with probability one to the true
joint distribution;
in other words, the empirical distribution gives a {\it consistent\/} estimate.
However, it is shown here that if the empirical distribution is used
to estimate the mixing coefficients,
then with probability one both the estimated $\beta$-mixing coefficient
and the estimated $\phi$-mixing coefficient approach one as $\lai$,
irrespective of what the true value might be.
Thus a quantity derived from a consistent estimator need not itself
be consistent.
\item On the other hand, if we bin the $l$ samples into $k_l$ bins
and choose $k_l$ in such a way that $k_l \ap \infty$ and
$k_l/l \ap 0$ as $\lai$, and a few technical conditions are satisfied,
then the empirically estimated $\al$-, $\beta$- and $\phi$-mixing
coefficients converge to their true values as $\lai$, with probability one.
\een

The notion of a mixing process and various definitions of mixing coefficients
originated in an attempt to establish the law
of large numbers for stationary stochastic processes that are not i.i.d.
The problem of determining (or at least bounding) the mixing coefficients
of random variables and stochastic processes arises in various contexts,
including system identification and statistical learning.
Traditional theories of system
identification are based on the assumption that the input sequence
to the unknown system is i.i.d.
However, it became clear over time that much of the theory continues to
hold even if the input sequence is not i.i.d., but is mixing in an
appropriate sense.
See \cite{Weyer00,VK07} as just two examples of such an approach.
Similarly, the standard formulation of PAC (probably approximately correct)
learning in statistical learning theory is based on the assumption
that the inputs are i.i.d.
See \cite{MV97,Vapnik98} for example.
However, subsequently PAC learning theory has been extended to the case
where the learning inputs are not i.i.d., but are mixing instead;
see for example the book \cite{MV03} and the references therein, as well
as \cite{Meir00}.
In adapting results in system identification or statistical 
learning theory from the i.i.d.\ case to the case of mixing processes,
it becomes necessary to obtain at least upper bounds for the mixing
coefficients, if not exact values.
The results presented here have some relevance to this problem, as do
other recent results such as \cite{MSS11}.
We shall return to this topic in the concluding remarks.

Proving that various mixing coefficients satisfy analogs of the data
processing inequality (DPI) is not just of academic interest.
Recent work on reverse-engineering genome-wide interaction networks
from gene expression data is based on first constructing a complete
graph where each node corresponds to a gene, and then using the DPI
to ``prune'' the graph.
Among the first algorithms to use this approach is ARACNE \cite{ARACNE},
which is based on using mutual information as a measure of interaction
between genes.
However, because mutual information is a symmetric quantity,
the resulting graphs are {\it undirected}, which is quite
contrary to biological realism, because in reality the interactions
between genes are not symmetric.
This led the authors to explore whether
the $\phi$-mixing coefficient, which is asymmetric, can be used
as a measure of the interaction between two genes.
Once it is established that the $\phi$-mixing coefficient satisfies an analog
of the DPI (which is one of the principal results of this paper),
it is possible to develop a method for constructing
{\it directed graphs\/} that represent whole genome regulatory networks.
However, this by itself is not sufficient.
If there are $n$ genes in the study, this approach requires the
computation of $n^2$ $\phi$-mixing coefficients.
So for a typical genome-wide study involving $20,000$ genes, it becomes
necessary to compute $400$ million $\phi$-mixing coefficients.
Hence it is mandatory to have a method for the
efficient computation of the $\phi$-mixing coefficient.
Such a method is also provided in the present paper.
Please see \cite{MV-EJC11,GIN,MV-Brief} for a discussion of how the
methods presented here can be applied to reverse engineering gene
regulatory networks.


\section{Definitions of Mixing Coefficients}\label{sec:2}

Definitions of the $\al$-, $\beta$- and $\phi$-mixing coefficients
of a stationary stochastic process can be found, among other places,
in \cite[pp.\ 34-35]{MV03}.
The $\al$-mixing coefficient was introduced by Rosenblatt \cite{Rosenblatt56}.
According to Doukhan \cite[p.\ 5]{Doukhan94}, Kolmogorov introduced
the $\beta$-mixing coefficient, but it appeared in print for the
first time in a paper published by some other authors.
The $\phi$-mixing coefficient was introduced by Ibragimov 
\cite{Ibragimov62}.

Essentially, all notions of mixing processes try to quantify the idea that,
in a stationary stochastic process of the form $\{ X_t \}_{t = -\infty}^\infty$,
the random variables $X_t$ and $X_\t$ become more and more independent
as $| t - \t |$ approaches infinity, in other words, there is an
asymptotic long-term near-independence.
However, these very general notions can be simplified and readily
adapted to define mixing coefficients between a pair of random
variables $X$ and $Y$.
This is how they are defined in \cite{Doukhan94}.
Note that, strictly speaking, mixing is a property not of
the random variables $X$ and $Y$, but rather of the $\s$-algebras
generated by $X$ and $Y$.
Note also that, if $\{ X_t \}_{t = -\infty}^\infty$ is a stationary stochastic
process, then the $k$-th ($\al$, $\beta$ or $\phi$) mixing coefficient of
the stochastic process is just the corresponding mixing coefficient
as defined in \cite{Doukhan94}
between the semi-infinite past $X_{-\infty}^0 := (X_t, t \leq 0)$
and the semi-infinite future $X_k^\infty := (X_t , t \geq k)$.

Though mixing coefficients
can be defined for arbitrary random variables, in the interests
of avoiding a lot of technicalities we restrict our attention in this paper to
just two practically important cases: real-valued and finite-valued random
variables.
We first define mixing coefficients between real-valued random variables,
and then between finite-valued random variables.

\begin{definition}\label{def:mix-real}
Suppose $X$ and $Y$ are real-valued random variables.
Let $\B$ denote the Borel $\s$-algebra of subsets of $\R$.
Then we define
\beq
\al(X,Y) & := & \sup_{S , T \in \B} | \Pr \{ X \in S \& Y \in T \} \nonumber \\
& - & \Pr \{ X \in S \} \cdot \Pr \{ Y \in T \} | .\label{eq:21}
\eeq
\beq
\phi(X|Y) & := & \sup_{S , T \in \B} 
| \Pr \{ X \in S | Y \in T \} - \Pr \{ X \in S \} | \nonumber \\
& = & \sup_{S , T \in \B}
\left| \frac{ \Pr \{ X \in S \& Y \in T \} } { \Pr \{ Y \in T \} }
- \Pr \{ X \in S \} \right| . \label{eq:22}
\eeq
\end{definition}

In applying the above definition, in case $\Pr \{ Y \in T \} = 0$,
we use the standard convention that
\bd
\Pr \{ X \in S | Y \in T \} = \Pr \{ X \in S \} .
\ed
Note that the $\al$-mixing coefficient is symmetric: $\al(X,Y) = \al(Y,X)$.
However, in general $\phi(X|Y) \neq \phi(Y|X)$.

The third coefficient, called the $\beta$-mixing coefficient, has
a somewhat more elaborate definition, at least in the general case.
Let $\th$ denote the probability measure of the joint random
variable $(X,Y)$, and let $\mu,\nu$ denote the marginal
measures of $X$ and $Y$ respectively.
Note that $\th$ is a measure on $\R^2$ while $\mu,\nu$ are measures
on $\R$.
If $X$ and $Y$ were independent, then $\th$ would equal $\mu \times
\nu$, the product measure.
With this in mind, we define
\be\label{eq:23}
\beta(X,Y) = \r(\th , \mu \times \nu) ,
\ee
where $\r$ denotes the total variation distance between two measures.
That is, if $\th,\eta$ are probability measures on a common measure space
$(\OM,\SI)$, then
\bd
\r(\th,\eta) := \sup_{S \in \SI} | \th(S) - \eta(S) | .
\ed
The $\beta$-mixing coefficient is also symmetric.

Next we deal with finite-valued random variables, and for this purpose we
introduce some notation that is used throughout the remainder of the paper.
The most important notational change is that, since probability
distributions on finite sets can be represented by vectors,
we use bold-face Greek letters
to denote them, whereas we use normal Greek letters to denote measures
on $\R$ or $\R^2$.
For each integer $n$, let $\Sm_n$ denote the $n$-dimensional simplex.
Thus
\bd
\Sm_n := \{ \v \in \R^n : v_i \geq 0 \fa i , \sum_{i=1}^n v_i = 1 \} .
\ed
If $\Abb = \{ a_1 , \ldots , a_n \}$ and $\bmu \in \Sm_n$, then
$\bmu$ defines a measure $P_\bmu$ on the set $\Abb$ according to
\bd
P_\bmu(S) = \sum_{i=1}^n \mu_i I_S(a_i) ,
\ed
where $I_S(\cdot)$ denotes the indicator function of $S$.

Suppose $\bmu,\bnu \in \Sm_n$ are probability distributions on
a set $\Abb$ of cardinality $n$.
Then the {\bf total variation distance} between $\bmu$ and $\bnu$
is defined as
\be\label{eq:24}
\r(\bmu,\bnu) := \max_{S \seq \Abb} | P_\bmu(S) - P_\bnu(S) | .
\ee
It is easy to give several equivalent closed-form formulas for the
total variation distance.
\bd
\r(\bmu,\bnu) = 0.5 \; \nm \bmu - \bnu \nm_1 
= \sum_{i=1}^n ( \mu_i - \nu_i )_+ 
= - \sum_{i=1}^n ( \mu_i - \nu_i )_- ,
\ed
where as usual $(\cdot)_+$ and $(\cdot)_-$ denote the nonnegative
and the nonpositive parts of a number:
\bd
(x)_+ = \max \{ x,0 \} , (x)_- = \min \{ x, 0 \} .
\ed

Now suppose $\Abb,\Bbb$ denote sets of cardinality $n,m$ respectively,
and that $\bmu \in \Sm_n , \bnu \in \Sm_m$.
Then the distribution $\bpsi \in \Sm_{nm}$ defined by 
$\psi_{ij} = \mu_i \nu_j$ is called the {\bf product distribution}
on $\Abb \times \Bbb$, and is denoted by $\bmu \times \bnu$.
In the other direction, if $\bth \in \Sm_{nm}$ is a distribution
on $\Abb \times \Bbb$, then $\bth_\Abb \in \Sm_n , \bth_\Bbb \in \Sm_m$
defined respectively by
\bd
(\bth_\Abb)_i := \sum_{j=1}^m \th_{ij} ,
(\bth_\Bbb)_j := \sum_{i=1}^n \th_{ij}
\ed
are called the {\bf marginal distributions} of $\bth$ on $\Abb$ and $\Bbb$
respectively.

The earlier definitions of mixing coefficients
become quite explicit in the case
where $X,Y$ are random variables assuming values in the finite
sets $\Abb,\Bbb$ of cardinalities $n,m$ respectively.
In this  case it does not matter whether the ranges of $X,Y$ are finite
subsets of $\R$ or some abstract finite sets.
Definition \ref{def:mix-real} can now be restated in this context.
Note that, since $\Abb,\Bbb$ are finite sets, the associated 
$\s$-algebras are just the power sets, that is, the collection of all subsets.

\begin{definition}\label{def:mix-finite}
With the above notation, we define
\be\label{eq:3a}
\al(X,Y) := \max_{S \seq \Abb , T \seq \Bbb}
| P_\bth ( S \times T ) - P_\bmu (S) P_\bnu(T) | ,
\ee
\be\label{eq:3b}
\beta(X,Y) := \r( \bth , \bmu \times \bnu ) ,
\ee
\be\label{eq:3c}
\phi(X|Y) := \max_{S \seq \Abb , T \seq \Bbb}
\left| \frac{ P_\bth( S \times T) }{ P_\bnu(T) } - P_\bmu(S) \right| .
\ee
\end{definition}

Whether $X,Y$ are real-valued or finite-valued random variables,
the mixing coefficients satisfy the following inequalities;
see \cite[p.\ 4]{Doukhan94}:
\be\label{eq:25}
\al(X,Y) \in [0,0.25] , \beta(X,Y) \in [0,1] , \phi(X,Y) \in [0,1] ,
\ee
\beq
0 & \leq & 2 \al(X,Y) \leq \beta(X,Y) \leq \min \{ \phi(X|Y) , \phi(Y|X) \}
\nonumber \\
& \leq & \max \{ \phi(X|Y) , \phi(Y|X) \} \leq 1 . \nonumber
\eeq
Also, the following statements are equivalent:
\ben
\item $X$ and $Y$ are independent random variables.
\item $\al(X,Y) = 0$.
\item $\beta(X,Y) = 0$.
\item $\phi(X|Y) = 0$.
\item $\phi(Y|X) = 0$.
\een

\section{Computation of Mixing Coefficients for Finite-Valued Random
Variables}\label{sec:3}


From the definitions, it is clear that $\beta(X,Y)$ can be readily
computed in closed form.
As before, let us define $\bpsi = \bmu \times \bnu$ to be the product
distribution of the two marginals, and define
\bd
\g_{ij} := \th_{ij} - \psi_{ij} , \GA := [ \g_{ij} ] \in [ -1, 1]^{n \times m} .
\ed
Then it readily follows from (\ref{eq:21}) that
\beq
\beta(X,Y) := \r( \bth , \bpsi ) 
& = & 0.5 \sum_{i=1}^n \sum_{j=1}^m | \g_{ij} | \nonumber \\
& = & \sum_{i=1}^n \sum_{j=1}^m ( \g_{ij} )_+ \nonumber \\
& = & - \sum_{i=1}^n \sum_{j=1}^m ( \g_{ij} )_- . \nonumber
\eeq
In addition, there is a very useful upper bound on the $\beta$-mixing
coefficient in terms of the so-called ``Pinsker's inequality'',
though it may be appropriate to credit this inequality to Csisz\'{a}r;
see \cite{CK81} or \cite{CT06}.
This inequality states that, for any two probability distributions
$\bth$ and $\bphi$ on a common set,
\bd
\r( \bth , \bphi ) \leq \sqrt { (1/2) D( \bth \nm \bphi ) } ,
\ed
where $D(\cdot,\cdot)$ is the Kullback-Leibler divergence.
Now apply this inequality with $\bphi = \bmu \times \bnu$.
This leads to
\bd
\r( \bth , \bmu \times \bnu) 
\leq \sqrt { (1/2) D( \bth \nm \bmu \times \bnu ) } ,
\ed
However, $\r( \bth , \bmu \times \bnu) = \beta(X,Y)$ whereas
$ D( \bth \nm \bmu \times \bnu ) = I(X,Y)$, the mutual information between
$X$ and $Y$.
Therefore
\bd
\beta(X,Y) \leq \sqrt{ (1/2) I(X,Y) } .
\ed

On the other hand, computing $\al(X,Y)$ or $\phi(X|Y)$ directly from
Definition \ref{def:mix-finite}
would require $2^{n+m}$ computations, since $S,T$ must
be allowed to vary over all subsets of $\Abb,\Bbb$ respectively.
Therefore the question arises as to whether this is an artefact
of the definition, or an inherent barrier to efficient computation.
In the present section, the following results are established:
\bit
\item As stated in (\ref{eq:25}), the quantity $\al(X,Y)$ always
lies in the interval $[0,0.25]$.
It is shown that the problem of determining whether $\al(X,Y) = 0.25$
for a given pair of random variables $X,Y$ is NP-complete.
More generally, given any number $\e \in (0,0.25]$, determining whether
$\al(X,Y) \geq \e$ is NP-complete.
\item Some efficiently computable upper and lower bounds are derived for
$\al(X,Y)$.
These bounds become germane in view of the above complexity result.
\item An exact and efficiently computable formula is derived for $\phi(X,Y)$.
\eit

In proceeding further, the first step is to
get rid of the absolute value signs in the definitions of the $\al$-
and $\phi$-mixing coefficients.

\begin{theorem}\label{thm:33}
It is the case that
\be\label{eq:34}
\al(X,Y) = \max_{S \seq \Abb , T \seq \Bbb}
\left[ P_\bth ( S \times T ) - P_\bmu (S) P_\bnu(T) \right] ,
\ee
\be\label{eq:35}
\phi(X|Y) = \max_{S \seq \Abb , T \seq \Bbb}
\left[ \frac{ P_\bth( S \times T) }{ P_\bnu(T) } - P_\bmu(S) \right] .
\ee
\end{theorem}

{\bf Proof:}
Define
\bd
{\cal R}_\al := \{ P_\bth ( S \times T ) - P_\bmu (S) P_\bnu(T) ,
S \seq \Abb , T \seq \Bbb \} .
\ed
Then ${\cal R}_\al$ is a finite subset of the real line consisting of at most
$2^{n+m}$ elements.
Now it is claimed that the set ${\cal R}_\al$ is symmetric; that is,
$x \in {\cal R}_\al$ implies that $-x \in {\cal R}_\al$.
If this claim can be established, then (\ref{eq:34}) follows readily.
So suppose $x \in {\cal R}_\al$, and choose $S \seq \Abb , T \seq \Bbb$
such that
\bd
P_\bth ( S \times T ) - P_\bmu (S) P_\bnu(T) = x .
\ed
Let $S^c$ denote the complement of $S$ in $\Abb$.
Then, using the facts that
\bd
P_\bmu(S^c) = 1 - P_\bmu(S) ,
\ed
\beq
P_\bth( S^c \times T) & = & P_\bth( \Abb \times T) - P_\bth ( S \times T ) 
\nonumber \\
& = & P_\bnu(T) - P_\bth ( S \times T ) , \nonumber
\eeq
it is easy to verify that
\bd
P_\bth(S^c \times T) - P_\bmu ( S^c) P_\bnu (T) = -x .
\ed
So ${\cal R}_\al$ is symmetric and (\ref{eq:34}) follows.
By analogous reasoning, the set
\bd
{\cal R}_\phi := \left\{ \frac{ P_\bth( S \times T) }{ P_\bnu(T) } - P_\bmu(S) :
S \seq \Abb , T \seq \Bbb \right\}
\ed
is also symmetric, which establishes (\ref{eq:35}).
$\halmos$

\subsection{NP-Completeness of Estimating the Alpha-Mixing Coefficient}\label{ssec:31}

We begin by revisiting the definition of $\al(X,Y)$, and determine 
the conditions under which it can attain its theoretical maximum
value of $0.25$.

\begin{theorem}\label{thm:3a}
Suppose $X,Y$ are random variables assuming values in finite sets $\Abb,\Bbb$
respectively, with marginal distributions $\bmu,\bnu$ respectively, and
joint distribution $\bth$.
Then $\al(X,Y) \leq 0.25$.
Moreover, $\al(X,Y) = 0.25$ if and only if there exist subsets
$S \seq \Abb,T \seq \Bbb$ such that $P_\bmu(S) = 0.5$,
$P_\bnu(T) = 0.5$, and $P_\bth(S \times T) = 0$.
\end{theorem}

{\bf Proof:} It is easy to see that the following relationship, which
is the mirror image of (\ref{eq:34}), is true:
\be\label{eq:3d}
\al(X,Y) = - \min_{S \seq \Abb , T \seq \Bbb}
\left[ P_\bth ( S \times T ) - P_\bmu (S) P_\bnu(T) \right] .
\ee
Indeed, as shown in the proof of Theorem \ref{thm:33}, if $S,T$ achieve
the maximum in (\ref{eq:34}), then $S^c,T$ achieve the minimum in
(\ref{eq:3d}), and vice versa.
Given sets $S \seq \Abb, T \seq \Bbb$, define
\bd
a := P_\bth(S \times T) , b := P_\bth(S \times T^c) ,
\ed
\bd
c := P_\bth(S^c \times T) , d := P_\bth(S^c \times T^c) .
\ed
Then it is evident that $a + b + c + d = 1$.
Moreover,
\bd
P_\bmu(S) = a + b , P_\bnu(T) = a + c .
\ed
Therefore
\beq
P_\bth ( S \times T ) - P_\bmu (S) P_\bnu(T) & = & a - (a+b) (a+c)
\nonumber \\
& = & -a^2 + a ( 1 - b - c ) - bc \nonumber \\
& =: & f(a) . \nonumber
\eeq
Let us think of the above quantity as a function of $a$ with $b,c$
fixed.
This amounts to fixing the measures of the sets $S,T$ while adjusting
the joint distribution to change $P_\bth(S \times T)$.
Then $f(0) = - bc$, and $f'(0) = 1 - b - c \geq 0$.
So $f(a)$ is nondecreasing at $a = 0$.
The maximum permissible value of $a$ is $1 - b - c$ (amounting to setting
$d = 0$), and $f(1 - b - c)$ again equals $-bc$.
Simple high school algebra shows that $f(a)$ achieves a maximum at
$a^* = (1 - b - c)/2$, and then begins to decrease.
Therefore it follows that $f(a) \geq - bc$.
Now $b,c$ satisfy $b+c \leq 1$, whence it is immediate that 
\beq
\al(X,Y) & \leq & - \min_{b,c} - bc \st b + c \leq 1 \nonumber \\
& = & \max_{b,c}  bc \st b + c \leq 1 \\ \nonumber
& = & 0.25 . \nonumber
\eeq
Moreover, $\al(X,Y) = 0.25$ if only if the choice $b = c = 0.5$
(which in turn implies that $a = d = 0$) is compatible with the
given joint distribution.
Recalling what these symbols represent shows that (i) $\al(X,Y) \leq 0.25$ 
always, and (ii) $\al(X,Y) = 0.25$ if and only if there exist subsets
$S \seq \Abb,T \seq \Bbb$ such that $P_\bth(S \times T) = 0$,
$P_\bmu(S) = P_\bnu(T) = 0.5$.
$\halmos$

The next step is to map a problem that is known to be NP-complete into
the problem of checking whether or not $\al(X,Y) = 0.25$.
Our choice is the so-called ``normalized partition'' problem, which
is a variant of the ``partition problem'', which 
can be found in \cite[p.\ 47]{Garey-Johnson79}, among other places.
We begin by stating the partition problem.

\noindent {\bf Problem Partition:}\\
{\bf Instance:} A positive integer $m$, and a set of positive integers
$a_1 , \ldots , a_m$.\\
{\bf Question:} Does there exist a subset $I \seq \M := \{ 1 , \ldots , m \}$
such that
\bd
\sum_{i \in I} a_i = \sum_{j \in \M \setminus I} a_j ?
\ed
This problem is known to be NP-Complete; see \cite[p.\ 47]{Garey-Johnson79}.
For our purposes we modify the problem as follows:

\noindent {\bf Problem Normalized Partition:}\\
{\bf Instance:} A positive integer $m$, and a set of positive rational
numbers $a_1 , \ldots , a_m$ such that $\sum_{i=1}^m a_i = 1$.\\
{\bf Question:} Does there exist a subset $I \seq \M := \{ 1 , \ldots , m \}$
such that
\bd
\sum_{i \in I} a_i = \sum_{j \in \M \setminus I} a_j ?
\ed
It is clear that this problem is equivalent to the partition problem,
and is therefore NP-complete.

\begin{theorem}\label{thm:3b}
The following problem is NP-complete:

\noindent {\bf Problem:}\\
{\bf Instance:} Positive integers $n,m$ and a set of nonnegative rational
numbers $\th_{ij}, i = 1 , \ldots , n , j = 1 , \ldots , m$ such that
$\sum_{i,j} \th_{ij} = 1$.\\
{\bf Question:} Let $(X,Y)$ be random variables assuming values
in $\N := \{ 1 , \ldots , n \}$, $\M := \{ 1 , \ldots , m \}$ respectively
with the joint distribution $\Pr \{ X = i \& Y = j \} =\th_{ij}$.
Is $\al(X,Y) = 0.25$?
\end{theorem}

{\bf Proof:} By Theorem \ref{thm:3a}, we know that $\al(X,Y) = 0.25$
if and only if there exist subsets $S \seq  \N, T \seq \M$, such that
$P_\bmu(S) = 0.5$, $P_\bnu(T) = 0.5$ and $P_\bth(S \times T) = 0$,
where $P_\bmu,P_\bnu$ denote the marginals of $P_\bth$.
Hence, given a candidate solution in terms of sets $S,T$, all one has to do
is to verify the above three relationships, which can be done in
polynomial time.
So the problem is in NP.

To show that it is NP-complete, we map the normalized partition problem
into it.
Given positive rational numbers $a_1 , \ldots , a_m$ such that
$\sum_{i=1}^m a_i = 1$, define $n = m$ and $\th_{ij} = a_i \d_{ij}$, where
$\d_{ij}$ is the Kronecker delta.
Thus, under this joint distribution, $n = m$, 
both $X$ and $Y$ have the vector
$\a = [ a_1 , \ldots , a_m]$ as their marginal distributions,
and $\Pr \{ X = Y \} = 1$.
Given subsets $S , T \seq \M$, it is easy to verify that
$P_\bth (S \times T) = P_\a(S \cap T)$.
(Note that $P_\bth$ is a measure on $\M \times \M$ while
$P_\a$ is a measure on $\M$.)
Therefore $\al(X,Y) = 0.25$ if and only if there exist subsets $S,T \seq
\M$ such that $P_\a(S) = 0.5$, $P_\a(T) = 0.5$, and $P_\a(S \cap T) = 0$.
These conditions imply that $S,T$ form a partition of $\M$, and that
$I = S, \M \setminus I = T$ is a solution of the normalized partition
problem. 
Hence this problem is NP-complete. $\halmos$

\begin{corollary}\label{corr:3a}
The following problem is NP-complete:

\noindent {\bf Problem:}\\
{\bf Instance:} Positive integers $n,m$, a set of nonnegative rational
integers $\th_{ij}, i = 1 , \ldots , n , j = 1 , \ldots , m$ such that
$\sum_{i,j} \th_{ij} = 1$, and a rational number $\e \in (0,0.25]$.\\
{\bf Question:} Let $(X,Y)$ be random variables assuming values
in $\N := \{ 1 , \ldots , n \}$, $\M := \{ 1 , \ldots , m \}$ respectively
with the joint distribution $\Pr \{ X = i \& Y = j \} =\th_{ij}$.
Is $\al(X,Y) \geq \e$?
\end{corollary}

{\bf Proof:}
If we choose $\e = 0.25$, this problem reduces to that studied in
Theorem \ref{thm:3b}, which is NP-complete.
Therefore the present problem is NP-hard.
On the other hand, given a candidate solution in terms of subsets
$S \seq \N,T \seq \M$, it is possible to verify in polynomial time
that $| P_\bth(S \times T) - P_\bmu(S) P_\bnu(T) | \geq \e$.
Therefore the problem is NP-complete. $\halmos$ 

\subsection{Upper and Lower Bounds for the Alpha-Mixing Coefficient}\label{ssec:32}

Since computing the $\al$-mixing coefficient is NP-hard (because merely
testing whether it exceeds a prespecified threshold is NP-complete),
it is worthwhile to have efficiently computable upper and lower bounds
for this mixing coefficient.
The aim of this subsection is to present such bounds.

To contrast with later results on the $\phi$-mixing coefficient,
we introduce a bit of notation.
Suppose $\GA \in \R^{n \times m}$, and that $p,q \in [1,\infty]$.
Then the induced norm $\nm \GA \nm_{p,q}$ is defined as
\bd
\nm \GA \nm_{p,q} := \max_{\nm \v \nm_p \leq 1} \nm \GA \v \nm_q .
\ed
Explicit closed-form formulas are available for $\nm \GA \nm_{p,q}$
when $ (p,q) = (1,1), (2,2), (\infty,\infty), (2,\infty)$;
see for example \cite{MV93}.
However, not much is known about other combinations.

\begin{theorem}\label{thm:3c}
Suppose $X,Y$ are random variables over finite sets $\Abb,\Bbb$ with
joint distribution $\bth$ and marginals $\bmu,\bnu$ respectively.
Define
\be\label{eq:3e}
\GA = \Th - \bmu \bnu^t \in \R^{n \times m},
\ee
where $\Th = [ \th_{ij} ]$.
Then
\be\label{eq:3f}
\al(X,Y) = 0.25 \; \nm \GA \nm_{\infty,1} .
\ee
\end{theorem}

{\bf Proof:}
Let $n$ denote $| \Abb |$, and
define a map $h : 2^\Abb \ap \bi^n$ as follows:
For a subset $S \seq \Abb$
\bd
h_i(S) = \left\{ \ba{lll}
1, & \mbox{if} & a_i \in S , \\
0, & \mbox{if} & a_i \not \in S .
\ea \right. \label{perm}
\ed
Note that by definition we have:
\bd
\h(S)+\h(S^c) = \eb_n,
\ed
where $\eb$ denotes a column vector whose components all equal one,
and the subscript denotes its dimension.
A similar map can be defined for $\Bbb$ as well.
With this notation, for any subsets $S \seq \Abb, T \seq \Bbb$, we have
that
\bd
P_\bmu(S) = \bmu^t \h(S) = [\h(S)]^t \bmu ,
\ed
\bd
P_\bnu(T) = \bnu^t \h(T) = [\h(T)]^t \bnu .
\ed
Moreover, with the joint distribution $\bth$, we have that
\bd
P_\bth(S \times T) = \bmu^t \Th \bnu .
\ed
Since the function $h$ is a bijection, it follows from (\ref{eq:34}) that
\beq
\al(X,Y) & = & \max_{ \a \in \{0,1\}^n, \b \in \{0,1\}^m }
\a^t \Th \b - \a^t \bmu \bnu^t \b \nonumber \\
\label{eq:3i}
& = & \max_{ \a \in \{0,1\}^n, \b \in \{0,1\}^m }  \a^t \GA \b .
\eeq

Let $\b \in \R^m$ be any fixed vector; then $\a^t \GA \b$ is
maximized with respect to $\a \in \bi^n$ by choosing $a_i = 1$
if $(\GA \b)i \geq 0$, and $a_i = 0$ if $(\GA \b)_i < 0$.
In other words,
\bd
\max_{\a \in \bi^n} \a^t \GA \b = \sum_i ( ( \GA \b )_i )_+ .
\ed
However, since the product distribution $\bmu \bnu^t$ and the joint
distribution $\Th$ have the same marginals, it follows that
\bd
\GA \eb_m = \bz_n , \eb_n^t \GA = \bz_m .
\ed
This implies that, for any vector $\a \in \bi^n$ and any $\b \in \R^m$, 
we have
\bd
\a^t \GA \b = - ( \eb_n - \a )^t \GA \b ,
\ed
and also that
\bd
\sum_i ( ( \GA \b )_i )_+ = - \sum_i ( ( \GA \b )_i )_- 
= 0.5 \; \nm \GA \b \nm_1 .
\ed
Therefore
\bd
\min_{\a \in \bi^n} \a^t \GA \b = - \max_{\a \in \bi^n} \a^t \GA \b ,
\ed
whence
\beq
\max_{\a \in \bi^n} | \a^t \GA \b | & = & \max_{\a \in \bi^n} \a^t \GA \b
\nonumber \\
& = & \sum_i ( ( \GA \b )_i )_+ \nonumber \\
& = & 0.5 \; \nm \GA \b \nm_1 . \nonumber
\eeq
As a consequence, it now follows from (\ref{eq:3i}) that
\beq
\al(X,Y) & = & \max_{\b \in \bi^m} \sum_i ( ( \GA \b )_i )_+ \nonumber \\
& = & 0.5 \; \max_{\b \in \bi^m} \nm \GA \b \nm_1 . \label{eq:3g}
\eeq

The proof is completed by showing that the quantity on the right side of
(\ref{eq:3g}) equals $\nm \GA \nm_{\infty,1}$.
For an arbitrary $\b \in \R^m$, define the associated vector $\z \in \R^m$ by
$\z = 2 \b - \eb_m$, and observe that, as $\b$ varies over
$\bi^m$, $\z$ varies over $\bp^m$.
Also,
\bd
\GA \b = 0.5 ( \GA \z - \GA \eb_m ) = 0.5 \GA \z ,
\ed
because $\GA \eb_m = \bz_m$.
Therefore
\bd
\al(X,Y) = 0.5 \; \max_{\b \in \bi^m} \nm \GA \b \nm_1
= 0.25 \; \max_{\z \in \bp^m} \nm \GA \z \nm_1 .
\ed
Now consider the optimization problem
\bd
\max \nm \GA \z \nm_\infty \st \nm \z \nm_\infty \leq 1 .
\ed
Since the objective function is convex and the feasible region is
convex and polyhedral, the optimum occurs at an extremum point.
In other words,
\bd
\max_{\nm \z \nm_\infty \leq 1} \nm \GA \z \nm_1 
= \max_{\z \in \bp^m} \nm \GA \z \nm_1 .
\ed
However, by definition the left side equals $\nm \GA \nm_{\infty,1}$.
This shows that
\bd
\al(X,Y) = 0.25 \; \nm \GA \nm_{\infty,1} ,
\ed
which is the desired conclusion. $\halmos$

By combining Theorems \ref{thm:3b} and \ref{thm:3c}, we can conclude
that computing the induced norm $\nm \cdot \nm_{\infty,1}$ of an arbitrary
matrix is NP-hard.
However, this result is already shown in \cite{Nesterov98}, which
also gives an efficiently computable upper bound for this induced norm,
with a guaranteed suboptimality.
By adapting that result, we can derive efficiently computable
upper and lower bounds for the $\al$-mixing coefficient.

\begin{theorem}\label{thm:3d}
\cite{Nesterov98}
Given a matrix $\GA \in \R^{n \times m}$, define $c(\GA)$ to be the
value of the following optimization problem:
\be\label{eq:3j}
c(\GA) := 0.5 \; \min_{\x \in \R^n , \y \in \R^m} 
\nm \x \nm _{\infty}+\nm \y \nm _1
\ee
subject to
\bd
\left[ \ba{cc}
{\rm Diag}( \x ) & \GA^t  \\
\GA &{\rm Diag}( \y )  \ea \right]
\geq 0 ,
\ed
where $M \geq 0$ denotes that $M$ is positive semidefinite.
Then
\be\label{eq:3k}
0.1086 \; c(\Gamma) \leq  \alpha(X,Y) \leq 0.25 \; c(\Gamma),
\ee
\end{theorem}

{\bf Proof:}
It is shown in \cite{Nesterov98} that
\bd
\nm \Gamma \nm _{\infty,1} \leq c(\Gamma) \leq 2.3 \; \nm \Gamma \nm _{\infty,1}.
\ed
Therefore
\bd
(1/2.3) \; c(\Gamma) \leq \nm \Gamma \nm _{\infty,1} \leq c(\Gamma) .
\ed
Combining this with (\ref{eq:3f}) leads to the desired conclusion. $\halmos$

Note that the computation of $c(\GA)$ requires the solution of
a semidefinite optimization program.
Efficient algorithms
to solve semidefinite programs can be found in \cite{VB96}.

\subsection{An Exact Formula for the Phi-Mixing Coefficient}\label{ssec:33}

We have seen in Section \ref{ssec:31} that estimating the $\al$-mixing
coefficient is NP-complete.
Though the definition of the $\phi$-mixing coefficient resembles
that of the $\al$-mixing coefficient in terms of (apparently) requiring
an enumeration of all subsets,
it turns out that there is an efficiently computable
exact formula for the $\phi$-mixing coefficient.

\begin{theorem}\label{thm:3e}
Suppose $X,Y$ are random variables over finite sets $\Abb,\Bbb$ with
joint distribution $\bth$ and marginals $\bmu,\bnu$ respectively.
Then
\beq
\phi(X,Y) & = & \max_j \frac{1}{\nu_j} \sum_i ( \g_{ij} )_+ \nonumber \\
& = & 0.5 \; \max_j \frac{1}{\nu_j} \sum_i | \g_{ij} | \nonumber \\
& = & 0.5 \; \nm \GA [ {\rm Diag} ( \bnu ) ]^{-1} \nm_{1,1} \label{eq:3h}
\eeq
\end{theorem}

{\bf Proof:}
We already know from Theorem \ref{thm:33} that
\bd
\phi(X|Y) := \max_{S \seq \Abb , T \seq \Bbb}
\Pr \{ X \in S | Y \in T \} - \Pr \{ X \in S \}  .
\ed
Now define
\bd
g(S,T) := \Pr \{ X \in S | Y \in T \} - \Pr \{ X \in S \}
\ed
Then
\bd
\phi(X|Y) = \max_{T \seq \Bbb} \max_{S \seq \Abb} g(s,T) .
\ed

Next, using obvious notation, let us rewrite $g(S,T)$ as
\beq
g(S,T) & = & P(S|T) - P(T) \nonumber \\
& = & \frac{ P(S \times T) - P(S) P(T) } { P(T) } .
\eeq
Now, suppose $T_1 , T_2$ are {\it disjoint\/} subsets of $\Bbb$.
Then
\bd
P(T _1 \cup T_2) = P(T_1) + P(T_2) ,
\ed
\bd
P(S \times (T_1 \cup T_2) ) = P(S \times T_1) + P(S \times T_2) 
\ed
because the events $S \times T_1$ and $S \times T_2$ are also disjoint.
Therefore
\beq
g(S,T_1 \cup T_2) & = &
\frac { P(S \times (T_1 \cup T_2) ) - P(T _1 \cup T_2) } { P(T _1 \cup T_2) }
\nonumber \\
& = & \frac{  P(S \times T_1 ) - P(T_1)  } { P(T_1) + P(T_2) } \nonumber \\
& + & \frac{ P(S \times T_2 ) - P(T_2)  }
{ P(T_1) + P(T_2) } \nonumber \\
& = & \frac{ P(S \times T_1 ) - P(T_1) }{ P(T_1) } 
\cdot \frac{ P(T_1) }{ P(T_1) + P(T_2) } \nonumber \\
& + & \frac{ P(S \times T_2 ) - P(T_2) }{ P(T_2) } 
\cdot \frac{ P(T_2) }{ P(T_1) + P(T_2) } \nonumber \\
& = & \l_1 g(S,T_1) + \l_2 g(S,T_2) , \nonumber
\eeq
where
\bd
\l_1 = \frac{ P(T_1) }{ P(T_1) + P(T_2) } ,
\l_2 = \frac{ P(T_2) }{ P(T_1) + P(T_2) } .
\ed
Therefore $g(S,T_1 \cup T_2)$ is a convex combination of
$g(S,T_1)$ and $g(S,T_2)$.
There is nothing special about writing $T$ as a disjoint union of
{\it two\/} subsets.
In general, if $T = \{ j_1 , \ldots , j_k \}$, then the above
reasoning can be repeated to show that
\bd
g(S,T) = \sum_{l=1}^k \frac{ P( \{ j_l \} ) }{ P(T) } g(S, \{ j_l \}) ,
\ed
that is, $g(S,T)$ is a convex combination of $g(S, \{ j_l \})$.
This shows that, if $T = \{ j_1 , \ldots , j_k \}$, then
\bd
g(S,T) \leq \max_{1 \leq l \leq k} g(S, \{ j_l \}) .
\ed
Hence, for a given subset $S \seq \Abb$, we have
\bd
\max_{T \seq \Bbb} g(S,T) = \max_{j \in \Bbb} g(S,\{ j \} ) .
\ed
The importance of the above equation lies in enabling us
to replace a maximum over all {\it subsets\/} of $\Bbb$ with
a maximum over all {\it elements\/} of $\Bbb$.
This is how we break through the barrier of enumerating an exponential
number of subsets.
As a consequence we have
\bd
\phi(X|Y) = \max_{j \in \Bbb} \max_{S \seq \Abb} g(S,\{ j \} ) .
\ed

Now for a fixed subset $S \seq \Abb$ and fixed element $j \in \Bbb$, we have
\beq
g(S,\{ j \} ) & = & \Pr \{ X \in S | Y = j \} - \Pr \{ X \in S \}
\nonumber \\
& = & \sum_{i \in S } \left[ \frac{ \th_{ij} }{ \nu_j } - \mu_i \right]
\nonumber \\
& = & \frac{1}{\nu_j} \sum_{i \in S } [ \th_{ij} - \mu_i \nu_j ]
\nonumber \\
& = & \frac{1}{\nu_j} \sum_{i \in S } \g_{ij} .
\nonumber
\eeq
Hence, for a fixed $j \in \Bbb$, the summation is maximized with respect
to $S$ by choosing $i \in S$ if $\g_{ij} \geq 0$ and $i \not \in S$
if $\g_{ij} < 0$.
The resulting maximum value (for a fixed $j \in \Bbb$) is
\bd
\max_{S \seq \Abb} g(S,\{ j \} ) = \frac{1}{\nu_j}
\sum_i (\g_{ij})_+ .
\ed
So finally
\beq
\phi(X|Y) & = & \max_{j \in \Bbb} \max_{S \seq \Abb} g(S,\{ j \} )
\nonumber \\
& = & \max_j \frac{1}{\nu_j} \sum_i (\g_{ij})_+ , \nonumber 
\eeq
which is the first equation in (\ref{eq:3h}).
The second equation in (\ref{eq:3h}) follows from the the fact that
$\eb_n^t \GA = \bz_m$, which implies in turn that, for each fixed $
j \in \Bbb$, we have that
\bd
\sum_i (\g_{ij})_+ = - \sum_i (\g_{ij})_- = 0.5 \sum_i |\g_{ij}| .
\ed
Lastly, the fact that
\bd
\max_j \frac{1}{\nu_j} 
\sum_i | \g_{ij} | = \nm \GA  [ {\rm Diag} ( \bnu ) ]^{-1} \nm_{1,1}
\ed
is standard and can be found in many places, e.g.\ \cite{MV93}. $\halmos$

We conclude this section by observing that the $\al$-mixing coefficient
is proportional to the $(\infty,1)$-induced norm of the matrix $\GA$,
whereas the
$\phi$-mixing coefficient is proportional to the $(1,1)$-induced
norm of the matrix $\GA  [ {\rm Diag} ( \bnu ) ]^{-1}$.
Therefore the reason for the NP-hardness of computing the $\al$-mixing 
coeffient and the efficient computability of the $\phi$-mixing coefficient
lies in the nature of the induced norms that need to be computed.

\section{Data Processing-Type Inequalities for Mixing Coefficients}\label{sec:6}

In this section we study the case where two finite-valued random variables are
conditionally independent given a third finite-valued random variable,
and prove inequalities of
the data processing-type for the associated mixing coefficients.
The nomenclature `data processing-type' is motivated by the well-known
data processing inequality in information theory.

\begin{definition}\label{def:cond}
Suppose $X,Y,Z$ are random variables assuming values in finite sets
$\Abb, \Bbb, \Cbb$ respectively.
Then $X,Z$ are said to be {\bf conditionally independent} given $Y$ if,
$\fa i \in \Abb , j \in \Bbb , k \in \Cbb$, it is true that
\beq
\Pr \{ X = i \& Z = k | Y = j \} & = & \Pr \{ X = i | Y = j \} \nonumber \\
& \times & \Pr \{ Z = k | Y = j \} .
\label{eq:61}
\eeq
\end{definition}

If $X,Z$ are conditionally independent given $Y$, we denote this by
$(X \perp Z) | Y$.
Some authors also write this as `$X \ap Y \ap Z$ is a short Markov chain',
ignoring the fact that the three random variables can belong to quite
distinct sets.
In this case, it makes no difference whether we write $X \ap Y \ap Z$
or $Z \ap Y \ap X$, because it is obvious from (\ref{eq:61}) that
conditional independence is a symmetric relationship.
Thus
\bd
(X \perp Z) | Y \iff (Z \perp X) | Y .
\ed
Also, from the definition, it follows readily that if $(X \perp Z) | Y$,
then $\fa S \seq \Abb , j \in \Bbb , U \seq \Cbb$ we have that
\beq
\Pr \{ X \in S \& Z \in U | Y = j \} & = &
\Pr \{ X \in S | Y = j \} \nonumber \\
& \times & \Pr \{ Z \in U | Y = j \} .\label{eq:62}
\eeq
However, in general, it is {\it not true\/} that,
$\fa S \seq \Abb ,T \seq \Bbb , U \seq \Cbb$,
\beq
\Pr \{ X \in S \& Z \in U | Y \in T \} & = &
\Pr \{ X \in S | Y \in T \} \nonumber \\
& \times & \Pr \{ Z \in U | Y \in T \} . \nonumber
\eeq
In fact, by setting $T = \Bbb$, it would follow from the above relationship that
$X$ and $Z$ are independent, which is a stronger requirement than conditional
independence.

Given two random variables $X,Y$ with joint distribution $\bth$ and
marginal distributions $\bmu,\bnu$ of $X,Y$ respectively, the
quantity
\bd
H( \bmu ) := - \sum_{i = 1}^n \mu_i \log \mu_i 
\ed
is called the {\bf entropy} of $\bmu$, with analogous definitions for
$H( \bnu)$ and $H( \bth )$; and the quantity
\bd
I(X,Y) = H( \bmu ) + H( \bnu ) - H( \bth) 
\ed
is called the {\bf mutual information} between $X$ and $Y$.
It is clear that $I(X,Y) = I(Y,X)$.
The following well-known inequality, referred to as the {\bf data-processing
inequality}, is the motivation for the contents of this section;
see \cite[p.\ 34]{CT06}.
Suppose $(X \perp Z) | Y$.
Then
\be\label{eq:63}
I(X,Z) \leq \min \{ I(X,Y) , I(Y,Z) \} .
\ee

\begin{theorem}\label{thm:61}
Suppose $(X \perp Z) | Y$.
Then
\be\label{eq:64}
\al(X,Z) \leq \min \{ \al(X,Y) , \al(Y,Z) \} .
\ee
\end{theorem}

\begin{theorem}\label{thm:62}
Suppose $(X \perp Z) | Y$.
Then
\be\label{eq:65}
\beta(X,Z) \leq \min \{ \beta(X,Y) , \beta(Y,Z) \} .
\ee
\end{theorem}

\begin{theorem}\label{thm:63}
Suppose $(X \perp Z) | Y$.
Then
\be\label{eq:66}
\phi(X|Z) \leq \min \{ \phi(X|Y) , \phi(Y|Z) \} ,
\ee
\be\label{eq:67}
\phi(Z|X) \leq \min \{ \phi(Z|Y) , \phi(Y|X) \} .
\ee
\end{theorem}

{\bf Proof of Theorem \ref{thm:61}:}
Let $S \seq \Abb , U \seq \Cbb$ be arbitrary, and define
\bd
r_\al(S,U) := \Pr \{ X \in S \& Z \in U \} - \Pr \{ X \in S \}
\Pr \{ Z \in U \} .
\ed
Then
\bd
\al(X,Y) = \max_{ S \seq \Abb , U \seq \Cbb} r_\al(S,U) .
\ed
Recall from (\ref{eq:3g}) that
\bd
\al(X,Y) = \max_{\b \in \bi^m} \sum_i ( ( \GA \b )_i )_+ .
\ed
Using the definition of the matrix $\GA$ and the one-to-one relationship
between vectors in $\bi^m$ and subsets of $\Bbb$, we can rewrite the
above equation equivalently as
\beq
\al(X,Y) = \max_{ T \seq \Bbb } \sum_{i=1}^n
& [ & \Pr \{ X = i \& Y \in T \} \nonumber \\
& - & \Pr \{ X = i \} \Pr \{ Y \in T \} ]_+ .\label{eq:312}
\eeq

Now we manipulate the quantity $r_\al(S,U)$ for arbitrary subsets $S \seq \Abb, 
U \seq \Cbb$ to prove the desired conclusion.\footnote{Due to the
width limitations of the two-column
format, the long equations that follow have been split across two
and sometimes three lines.}
\beq
r_\al(S,U) & = & \sum_{j=1}^m [ \Pr \{ X \in S \& Y = j \& Z \in U \}
\nonumber \\
& - & \Pr \{ X \in S \& Y = j \} \Pr \{ Z \in U \} ] \nonumber \\
& = & \sum_{j=1}^m [ \Pr \{ X \in S | Y = j \} \Pr \{ Z \in U | Y = j \} 
\nonumber \\
& \times & \Pr \{ Y = j \} \nonumber \\
& - & \Pr \{ X \in S | Y = j \} \Pr \{ Y = j \} \Pr \{ Z \in U \} ] \nonumber \\
& = & \sum_{j=1}^m \Pr \{ X \in S | Y = j \} 
[ \Pr \{ Z \in U \& Y = j \} 
\nonumber \\
& - & \Pr \{ Y = j \} \Pr \{ Z \in U \} ] \nonumber \\
& \leq & \sum_{j=1}^m \Pr \{ X \in S | Y = j \}
[ \Pr \{ Z \in U \& Y = j \} 
\nonumber \\
& - & \Pr \{ Y = j \} \Pr \{ Z \in U \} ]_+
\nonumber \\
& \leq & \sum_{j=1}^m
[ \Pr \{ Z \in U \& Y = j \} 
\nonumber \\
& - & \Pr \{ Y = j \} \Pr \{ Z \in U \} ]_+ 
\nonumber \\
& \leq & \max_{U \seq \Cbb} \sum_{j=1}^m 
[ \Pr \{ Z \in U \& Y = j \} 
\nonumber \\
& - & \Pr \{ Y = j \} \Pr \{ Z \in U \} ]_+
\nonumber \\
& = & \al(Y,Z) . \nonumber
\eeq

Since $S$ and $U$ are arbitrary, this implies that $\al(X,Z) \leq \al(Y,Z)$
whenever $X \ap Y \ap Z$ is a short Markov chain.
Since $X \ap Y \ap Z$ is the same as $Z \ap Y \ap X$, it also follows
that $\al(Z,X) \leq \al(Y,X)$.
Finally, since $\al$ is symmetric, the desired conclusion (\ref{eq:64}) 
follows.
$\halmos$

{\bf Proof of Theorem \ref{thm:62}:}
Suppose that $\Abb,\Bbb,\Cbb$ have cardinalities $n,m,l$ respectively.
(The symbols $n,m$ have been introduced earlier and now $l$ is introduced.)
Let $\bdelta$ denote the joint distribution of $(X,Y,Z)$, $\bzeta$
the joint distribution of $(X,Z)$, $\boldeta$ the joint distribution of
$(Y,Z)$, and as before, $\bth$ the joint distribution of $(X,Y)$.
Let $\bxi$ the marginal distribution of $Z$, and as before, let
$\bmu, \bnu$ denote the marginal distributions of $X$ and $Y$.
Finally, define 
\bd
c_{jk} = \frac{ \eta_{jk} }{ \nu_j } = \Pr \{ Z = k | Y = j \} .
\ed
As can be easily verified,
the fact that $(X \perp Z ) | Y$ (or (\ref{eq:61})) is equivalent to
\bd
\d_{ijk} = \frac{ \th_{ij} \eta_{jk} }{ \nu_j } 
= \th_{ij} c_{jk} , \fa i , j , k .
\ed
Also note the following identities:
\bd
\sum_{i=1}^n \th_{ij} = \nu_j ,
\sum_{j=1}^m \th_{ij} = \mu_i ,
\sum_{j=1}^m \d_{ijk} = \zeta_{ik} ,
\fa i, j, k .
\ed
Now it follows from the various definitions that
\beq
\beta(X,Z) & = & \sum_{i=1}^n \sum_{k=1}^l ( \zeta_{ik} - \mu_i \xi_k ) _+
\nonumber \\
& = & \sum_{i=1}^n \sum_{k=1}^l \left(
\sum_{j=1}^m ( \d_{ijk} - \th_{ij} \xi_k ) \right)_+ \nonumber \\
& \leq & \sum_{i=1}^n \sum_{k=1}^l \sum_{j=1}^m 
( \d_{ijk} - \th_{ij} \xi_k )_+ \nonumber \\
& = & \sum_{i=1}^n \sum_{k=1}^l \sum_{j=1}^m 
\left( \th_{ij} c_{jk} - \th_{ij} \xi_k \right)_+ \nonumber \\
& = & \sum_{k=1}^l \sum_{j=1}^m
\left[ \sum_{i=1}^n \th_{ij} \right] ( c_{jk} - \xi_k )_+ \nonumber \\
& = & \sum_{k=1}^l \sum_{j=1}^m ( \nu_j c_{jk} - \nu_j \xi_k )_+ \nonumber \\
& = & \sum_{k=1}^l \sum_{j=1}^m ( \eta_{jk} - \nu_j \xi_k )_+ \nonumber \\
& = & \beta(Y,Z) . \nonumber
\eeq
Now the symmetry of $\beta(\cdot,\cdot)$ serves to show that
$\beta(X,Z) \leq \beta(X,Y)$.
Putting both inequalities together leads to the desired conclusion.
$\halmos$

{\bf Proof of Theorem \ref{thm:63}:}
Suppose $(X \perp Z) | Y$.
Since the $\phi$-mixing coefficient is not symmetric, it is necessary to
prove two distinct inequalities, namely:
(i) $\phi(X|Z) \leq \phi(X|Y)$, and (ii) $\phi(X|Z) \leq \phi(Y|Z)$.

{\bf Proof that $\phi(X|Z) \leq \phi(X|Y)$:}
For $S \seq \Abb$, define
\bd
r_\phi(S) := \max_{T \seq \Bbb} \Pr \{ X \in S | Y \in T \} ,
\ed
and observe that
\bd
\phi(X|Y) = \max_{S \seq \Abb} [ r_\phi(S) - P_\bmu(S) ] .
\ed
Suppose $S \seq \Abb, U \seq \Cbb$ are arbitrary.
Then
\beq
\Pr \{ X \in S \& Z \in U \}
& = & \sum_{j=1}^m \Pr \{ X \in S \& Y = j \& Z \in U \} \nonumber \\
& = & \sum_{j=1}^m \Pr \{ X \in S | Y = j \}
\nonumber \\
& \times &
\Pr \{ Z \in U | Y = j \} \Pr \{ Y = j \} \nonumber \\
& = & \sum_{j=1}^m \Pr \{ X \in S | Y = j \}
\nonumber \\
& \times &
\Pr \{ Z \in U \& Y = j \} \nonumber \\
& \leq & r_\phi(S) \sum_{j=1}^m \Pr \{ Z \in U \& Y = j \} \nonumber \\
& = & r_\phi(S) \Pr \{ Z \in U \} . \nonumber
\eeq
Dividing both sides by $\Pr \{ Z \in U \}$ leads to
\bd
\Pr \{ X \in S | Z \in U \} \leq r_\phi(S) ,
\ed
\bd
\Pr \{ X \in S | Z \in U \} - P_\bmu(S) \leq r_\phi(S) - P_\bmu(S) 
\ed
Taking the maximum of both sides with respect to $S \seq \Abb$,
$U \seq \Cbb$ shows that
\bd
\phi(X|Z) \leq \phi(X|Y) .
\ed

{\bf Proof that $\phi(X|Z) \leq \phi(Y|Z)$:}
We begin by rewriting the expression for $\phi(X|Y)$.
In order to make the equations fit, for $S \seq \Abb, T \seq \Bbb$
we will use $P(S|T)$ as a shorthand for $\Pr \{ X \in S | Y \in T \}$, 
and so on.
With this convention, for $S \seq \Abb, T \seq \Bbb$, we have that
\beq
P(S|T) - P(S) & = & \sum_{i \in S} [ P(i|T) - P(i) ] \nonumber \\
& \leq & \sum_{i \in S} [ P(i|T) - P(i) ]_+ \nonumber \\
& \leq & \sum_{i \in \Abb} [ P(i|T) - P(i) ]_+ . \nonumber
\eeq
Therefore
\beq
\phi(X|Y) & = & \max_{S,T} [ P(S|T) - P(S) ] \nonumber \\
& \leq & \max_T \sum_{i \in \Abb} [ P(i|T) - P(i) ]_+ . \label{eq:3l}
\eeq
Actually this can be shown to be an equality, and not an inequality,
but we will not expend space on that.

Let us define
\bd
c(S,U) := \Pr \{ X \in S | Z \in U \} - P_\bmu(S) ,
\ed
and reason as follows:
\beq
c(S,U) & = & \Pr \{ X \in S | Z \in U \} - \Pr \{ X \in S \} \nonumber \\
& = & \sum_{j=1}^m 
[ \Pr \{ X \in S \& Y = j | Z \in U \} 
\nonumber \\
& - & \Pr \{ X \in S \& Y = j \} ] 
\nonumber \\
& = & \sum_{j=1}^m
[ \Pr \{ X \in S | Y = j \& Z \in U \} 
\nonumber \\
& \times & \Pr \{ Y = j | Z \in U \} 
\nonumber \\
& - & \Pr \{ X \in S | Y = j \} \Pr \{ Y = j \} ] 
\nonumber \\
& = & \sum_{j=1}^m
\Pr \{ X \in S | Y = j \} 
[ \Pr \{ Y = j | Z \in U \} 
\nonumber \\
& - & \Pr \{ Y = j \} ] \nonumber \\
& \leq & \sum_{j=1}^m
\Pr \{ X \in S | Y = j \} 
[ \Pr \{ Y = j | Z \in U \} 
\nonumber \\
& - & \Pr \{ Y = j \} ]_+ \nonumber \\
& \leq & \sum_{j=1}^m
[ \Pr \{ Y = j | Z \in U \} - \Pr \{ Y = j \} ]_+ \nonumber \\
& \leq & \max_{U \seq \Cbb} \sum_{j=1}^m
[ \Pr \{ Y = j | Z \in U \} - \Pr \{ Y = j \} ]_+ \nonumber \\
& \leq & \phi(Y|Z) , \nonumber
\eeq
where the last step follows from (\ref{eq:3l}).
Since the right side is independent of both $S$ and $U$, the desired
conclusion follows.
$\halmos$.


\section{Inconsistency of an Estimator for Mixing Coefficients}\label{sec:4}

Suppose $X,Y$ are real-valued random variables with some unknown
joint distribution,
and suppose we are given an infinite sequence of independent
samples $\{ (x_i, y_i) , i = 1 , 2 , \ldots \}$.
The question studied in this section and the next is whether it is possible
to construct empirical estimates of the various mixing coefficients
that converge to the true values as the number of samples approaches infinity.

Let
\bd
\Phi_{X,Y}(a,b) = \Pr \{ X \leq a \& Y \leq b \} 
\ed
denote the true but unknown joint distribution function of $X$ and $Y$, and
let $\Phi_X(\cdot),\Phi_Y(\cdot)$ denote the true but unknown
marginal distribution functions of $X,Y$ respectively.
Using the samples$\{ (x_i, y_i) , i = 1 , 2 , \ldots \}$,
we can construct three `stair-case functions'
that are empirical estimates of $\Phi_X,\Phi_Y$ and $\Phi_{X,Y}$
based on the first $l$ samples, as follows:
\be\label{eq:41}
\Phih_X(a;l) := \frac{1}{l} \sum_{i=1}^l I_{ \{ x_i \leq a \} } ,
\ee
\be\label{eq:42}
\Phih_Y(b;l) := \frac{1}{l} \sum_{i=1}^l I_{ \{ y_i \leq b \} } ,
\ee
\be\label{eq:43}
\Phih_{X,Y}(a,b;l) := \frac{1}{l} \sum_{i=1}^l I_{ \{ x_i \leq a 
\& y_i \leq b \} } ,
\ee
where as usual $I$ denotes the indicator function.
Thus $\Phih_X(a;l)$ counts the fraction of the first $l$ samples
that are less than or equal to $a$, and so on.
With this construction, the well-known Glivenko-Cantelli
lemma (see \cite{Glivenko33,Cantelli33} or \cite[p.\ 20]{Loeve77})
states that the empirical estimates converge uniformly and
almost surely to their true functions as the number of samples $\lai$.
Thus $\Phih_{X,Y}$ is a consistent estimator of the true joint distribution.
Thus one might be tempted to think that an empirical estimate of
any (or all) of the three mixing coefficients based on 
$\Phih_{X,Y}$ will also converge to the true
value as $\lai$.
The objective of this brief section is to show that this is not so.
Hence estimates of mixing coefficients derived from a consistent estimator
of the joint distribution need not themselves be consistent.

\begin{theorem}\label{thm:41}
Suppose $\Phih_{X,Y}$ is defined by (\ref{eq:43}), and that
$x_i \neq x_j$ and $y_i \neq y_j$ whenever $i \neq j$.
Let $\betah_l$ denote the $\beta$-mixing coefficient associated with
the joint distribution $\Phih_{X,Y}(\cdot,\cdot;l)$.
Then $\betah_l = (l-1)/l$.
\end{theorem}

{\bf Proof:}
Fix the integer $l$ in what follows.
Note that the empirical distribution $\Phih_{X,Y}(\cdot,\cdot;l)$
depends only the totality of the $l$ samples, and not the order
in which they are generated.
Without loss of generality, we can replace the samples $(x_1 , \ldots , x_n)$
by their `order statistics', that is, the same samples arranged in
increasing order, and do the same for the $y_i$.
Thus the assumption is that $x_1 < x_2 < \ldots < x_l$ and
similarly $y_1 < y_2 < \ldots < y_l$.
With this convention, the empirical samples will be of the form
$\{ (x_1 , y_{\pi(1)}) , \ldots , (x_l , y_{\pi(l)}) \}$ for some
permutation $\pi$ of $\{ 1, \ldots , l \}$.
Therefore the probability measure associated with the empirical
distribution $\Phih$ is purely atomic, with jumps of magnitude $1/l$
at the points $\{ (x_1 , y_{\pi(1)}) , \ldots , (x_l , y_{\pi(l)}) \}$.
So we can simplify matters by replacing the real line on the $X$-axis
by the finite set $\{ x_1 , \ldots , x_l \}$, and the real line on the
$Y$-axis by the finite set $\{ y_1 , \ldots , y_l \}$.
With this redefinition, the joint distribution $\bth$ assigns a weight
of $1/l$ to each of the points $(x_i, y_{\pi(i)})$ and a weight of zero
to all other points $(x_i , y_j)$ whenever $j \neq \pi(i)$, while the
marginal measures $\bmu,\bnu$ of $X$ and $Y$ will be uniform
on the respective finite sets.
Thus the product measure $\bmu \times \bnu$ assigns a weight of $1/l^2$
to each of the $l^2$ grid points $(x_i,y_j)$.
From this, it is easy to see that
\bd
\betah_l = \r(\bth,\bmu \times \bnu) = (l-1)/l .
\ed
This is the desired conclusion.
$\halmos$

\begin{corollary}\label{corr:41}
Suppose the true but unknown distribution $\Phi_{X,Y}$ has a density
with respect to the Lebesgue measure.
Then $\betah_l \ap 1, \phih_l \ap 1$ almost surely as $\lai$.
\end{corollary}

{\bf Proof:}
If the true distribution has a density, then it is nonatomic, which
means that with probability one, samples will be pairwise distinct.
It now follows from Theorem \ref{thm:41} that
\bd
\phih_l \geq \betah_l = \frac{l-1}{l} \ap 1 \mbox{ as } \lai .
\ed
This is the desired conclusion.
$\halmos$

\section{Consistent Estimators for Mixing Coefficients}\label{sec:5}

The objective of the present section is to show that a simple
modification of the `naive' algorithm proposed in Section \ref{sec:4}
does indeed lead to consistent estimates, provided appropriate technical
conditions are satisfied.

The basic idea behind the estimators is quite simple.
Suppose that 
one is given samples $\{ (x_i , y_i) , i \geq 1 \}$ generated independently
and at random from an unknown joint probability measure $\th \in \M(\R^2)$.
Given $l$ samples, choose an integer $k_l$ of bins.
Divide the real line into $k_l$ intervals such that each bin contains
$\lfloor l/k_l \rfloor$ or $\lfloor l/k_l \rfloor + 1$ samples for both
$X$ and $Y$.
In other words, carry out percentile binning of both random variables.
One way to do this (but the proof is not dependent on how precisely this
is done) is as follows:
Define $m_l = \lfloor l/k_l \rfloor , r = l - k_l m_l$, and place
$m_l +1$ samples in the first $r$ bins and $m_l$ samples in the next $m_l - r$
bins.
This gives a way of discretizing the real line for both $X$ and $Y$
such that the discretized random variables have nearly uniform
marginals.
With this binning, compute the corresponding joint distribution,
and the associated empirical estimates of the mixing coefficients.
The various theorems below show that, subject to some regularity
conditions, the empirical estimates produced by this scheme do indeed
converge to their right values with probability one as $\lai$,
{\it provided that\/} $m_l \ap \infty$, or equivalently, $k_l / l \ap 0$,
as $\lai$.
In other words, in order for this theorem to apply,
the number of bins must increase more slowly than the
number of samples, so that the number of samples per bin must approach
infinity.
In contrast, in Theorem \ref{thm:41}, we have effectively chosen $k_l = l$
so that each bin contains precisely one sample, which explains why
that approximation scheme does not work.

To state the various theorems,
we introduce a little bit of notation, and refer
the reader to \cite{Berberian70} for all concepts from measure theory
that are not explicitly defined here.
Let $\M(\R), \M(\R^2)$ denote the set of all measures on $\R$ or $\R^2$
equipped with the Borel $\s$-algebra.
Recall that if $\th , \eta \in \M(\R)$ or $\M(\R^2)$, then $\th$ is
said to be {\bf absolutely continuous} with respect to $\eta$, denoted
by $\th \ll \eta$, if for every measurable set $E$,
$\eta(E) = 0 \imp \th(E) = 0$.

Next, let $\th$ denote the joint probability measure of $(X,Y)$,
and let $\mu,\nu$ denote the marginal measures.
Thus, for every measurable\footnote{Hereafter we drop this adjective; it
is assumed that all sets that are encountered are measurable.}
subset $S \seq \R$, the measure $\mu(S)$ is defined as $\th(S \times \R)$\
and similarly for all $T \seq \R$, the measure $\nu(T)$ is defined
as $\th(\R \times T)$.
Now the key assumption made here is that {\it the joint measure $\th$
is absolutely continuous with respect to the product measure $\mu \times \nu$}.
In the case of finite-valued random variables, this assumption is automatically
satisfied.
Suppose that for some pair of indices $i,j$, it is the case that
$\mu_i \cdot \nu_j = 0$.
Then either $\mu_i = 0$ or $\nu_j = 0$.
If $\mu_i = 0$, then it follows from the identity 
$\sum_{j'} \th_{i j'} = \mu_i$ that $\th_{ij'} = 0$ for all $j'$,
and in particular $\th_{ij} = 0$.
Similarly if $\nu_j = 0$, then it follows from the identity
$\sum_{i'} \th_{i'j} = \nu_j$ that $\th_{i'j} = 0$ for all $i'$,
and in particular $\th_{ij} = 0$.
In either case it follows that $\th_{ij} = 0$, so that $\th \ll \mu \times \nu$.
However, in the case of real random variables, this need not be so.
For example, replace $\R \times \R$ by the unit square, and let $\th$
be the diagonal measure.
Then both marginals $\mu,\nu$ are the uniform measures on the unit
interval, and the product $\mu \times \nu$ is the uniform measure
on the unit square -- and $\th$ is singular with respect to the
uniform measure.

Next we introduce symbols for the various densities.
Since $\th \ll \mu \times \nu$, it follows that $\th$ has a 
Radon-Nikodym derivative
with respect to $\mu \times \nu$, which is denoted by $f(\cdot,\cdot)$.
So for any sets $S, T \seq \R$, it follows that
\beq
\th(S \times T) & = & \int_S \int_T f(x,y) d \nu(y) d \mu(x)
\nonumber \\
& = & \int_T \int_S f(x,y) d \mu(x) d \nu(y) .
\nonumber
\eeq
For any $T \seq \R$ with $\nu(T) > 0$, the conditional probability
$\Pr \{ X \in S | Y \in T \}$ is given by
\beq
\Pr \{ X \in S | Y \in T \} & = & \frac{ \Pr \{ X \in S \& Y \in T \} }
{ \Pr \{ Y \in T \} } = \frac{ \th(S \times T) }{ \nu(T) }
\nonumber \\
& = & \int_S \left[ \int_T \frac{ f(x,y) }{ \nu(T) } d \nu(y) \right] d \mu(x) .
\nonumber
\eeq

\begin{theorem}\label{thm:51}
Suppose $\th \ll \mu \times \nu$,
and that $k_l \ap \infty$, $k_l/l \ap 0$ as $\lai$.
Then the empirically estimated
$\beta$-mixing coefficient $\betah_l$ converges almost surely to
the true value $\beta$ as $\lai$.
\end{theorem}

\begin{theorem}\label{thm:52}
Suppose $\th \ll \mu \times \nu$,
and in addition that the density $f(\cdot,\cdot)$ belongs to
$L_\infty(\R^2)$.
Suppose that $k_l \ap \infty$, $k_l/l \ap 0$ as $\lai$.
Then  the empirically estimated
$\al$-mixing coefficient $\alh_l$ converges almost surely to
the true value $\al$ as $\lai$, and
the empirically estimated
$\phi$-mixing coefficient $\phi_l$ converges almost surely to
the true value $\phi$ as $\lai$.
\end{theorem}


Note that the absolute continuity assumption $\th \ll \mu \times \nu$
guarantees that the density $f(\cdot,\cdot) \in L_1(\R^2, \mu \times \nu)$.
So no additional technical assumptions are needed to ensure that
the sequence of empirical estimates $\betah_l$ converges to its true value.
However, in order to establish that 
the sequences of empirical estimates $\alh_l$ and $\phih_l$
converge to their true values, we have added an assumption that
the density $f$ is bounded almost everywhere.
This condition
is intended to ensure
that conditional densities do not `blow up'.
In the case of finite-valued variables,
we have already seen that the condition $\th \ll \mu \times \nu$ holds
automatically, which means that the `density' $f_{ij} \th_{ij}/(\mu_i \nu_j)$
is always well-defined.
Since there are only finitely many values of $i$ and $j$, this
ratio is also bounded.
However, in the case of real-valued random variables, this condition
needs to be imposed explicitly.

The proofs of these two theorems are based on arguments in \cite{LN96,WKV05}.
In the proof of Theorem \ref{thm:51}, we can use those arguments as they are,
whereas in the proof of Theorem \ref{thm:52}, we need to adapt them.
To facilitate the discussion, we first reprise the relevant results
from \cite{LN96,WKV05}.

\begin{definition}\label{def:51}
Let $( \OM, \F )$ be a measurable space, and let $Q$ be a probability
measure on $( \OM, \F )$.
Suppose $\{ I_1 , \ldots , I_L \}$ is a finite
partition of $\OM$, and that $\{ I_1^{(m)} , \ldots , I_L^{(m)} \}$
is a sequence of partitions of $\OM$.
Then $\{ I_1^{(m)} , \ldots , I_L^{(m)} \}$ is said to {\bf converge to}
$\{ I_1 , \ldots , I_L \}$ with respect to $Q$ if, for every probability measure
$P$ on $( \OM, \F )$ such that $P \ll Q$, it is the case that
\bd
P(I_i^{(m)}) \ap P(I_i) \mbox{ as } \mai .
\ed
\end{definition}
See \cite[Definition 1]{WKV05}.

\begin{theorem}\label{thm:54}
Suppose $Q$ is a probability measure on $(\R,\B)$ that is absolutely
continuous with respect to the Lebesgue measure, $L$ is a fixed integer,
and that $\{ I_1 , \ldots , I_L \}$ is an equiprobable partitioning of $\R$.
In other words, choose numbers
\bd
- \infty = a_0 < a_1 < \ldots < a_{L-1} < a_L = + \infty 
\ed
such that the semi-open intervals $I_i = (a_{i-1},a_i]$ satisfy
\bd
Q(I_i) = 1/L , i = 1 , \ldots , L .
\ed
Suppose $\{ y_1 , \ldots , y_m \}$ are i.i.d.\ samples generated in
accordance with $Q$, and that $m = l_m T$ with $l_m \in \Nbb$, an integer.
Let $\{ I_1^{(m)} , \ldots , I_L^{(m)} \}$ denote the empirical
equiprobable partitioning associated with the samples
$\{ y_1 , \ldots , y_m \}$.
Then $\{ I_1^{(m)} , \ldots , I_L^{(m)} \}$ converges to
$\{ I_1 , \ldots , I_L \}$ with respect to $Q$ as $\mai$.
\end{theorem}

{\bf Proof:} See \cite[Lemma 1]{WKV05}.

\begin{theorem}\label{thm:55}
Let $( \OM, \F )$ be a measurable space, and let $Q$ be a probability
measure on $( \OM, \F )$.
Suppose $\{ I_1^{(m)} , \ldots , I_L^{(m)} \}$
is a sequence of partitions of $\OM$ that converges with respect to $Q$ to
another partition $\{ I_1 , \ldots , I_L \}$ as $\mai$.
Suppose $\{ x_1 , \ldots , x_n \}$ are i.i.d.\ samples generated in
accordance with a probability measure $P \ll Q$, and let $P_n$
the empirical measure generated by these samples.
Then
\bd
\lim_{\mai} \lim_{\nai} P_n(I_i^{(m)}) = P(I_i) , \mbox{ a.s. } \fa i .
\ed
\end{theorem}

{\bf Proof:} See \cite[Lemma 2]{WKV05}.

Before proceeding to the proofs of the two theorems, we express
the three mixing coefficients in terms of the densities.
As stated in (\ref{eq:23}), we have that
\be\label{eq:43a}
\beta(X,Y) = 0.5 \int_\R \int_\R | f(x,y) - 1 | d \mu (x) d \nu (y) .
\ee
Here we take advantage of the fact that the `density' of $\mu$
with respect to itself is one, and similarly for $\nu$.
Next, as in Theorem \ref{thm:33}, we can drop the absolute value signs in
the definitions of $\al(X,Y)$ and of $\phi(X|Y)$.
Therefore the various mixing coefficients can be expressed as follows:
\be\label{eq:44}
\al(X,Y) = \sup_T \sup_S
\int_S \int_T [ f(x,y) - 1 ] d \nu (y) d \mu (x) ,
\ee
\be\label{eq:45}
\phi(X,Y) =  \sup_T \sup_S 
\int_S \left[ \int_T \frac{ f(x,y) }{ \nu(T) } d \nu (y) 
- 1 \right] d \mu (x) .
\ee
Now, for each fixed set $T$, let us define signed measures
$\kappa_T$ and $\d_T$ as follows:
\bd
\kappa_T(x) = \int_T [ f(x,y) - 1 ] d \nu (y) ,
\ed
\bd
\d_T(x) = \int_T \frac{ f(x,y) }{ \nu(T) } d \nu (y) - 1 ,
\ed
and associated support sets
\bd
A_+(T) = \{ x \in \R : \kappa_T(x) \geq 0 \} ,
\ed
\bd
B_+(T)  = \{ x \in \R : \d_T(x) \geq 0 \} .
\ed
Then it is easy to see that, for each fixed set $T$, the supremum
in (\ref{eq:44}) is achieved by the choice $S = A_+(T)$ while
the supremum in (\ref{eq:45}) is achieved by the choice $S = B_+(T)$.
Therefore
\beq
\al(X,Y)
& = & \sup_T \int_{A_+(T)} \kappa_T(x) d \mu (x) 
\nonumber \\
& = & \sup_T \int_\R [ \kappa_T(x) ]_+ d \mu (x) ,
\label{eq:46}
\eeq
\beq
\phi(X|Y)
& = & \sup_T \int_{B_+(T)} \d_T(x) d \mu (x) 
\nonumber \\
& = & \sup_T \int_\R [ \d_T(x) ]_+ d \mu (x) .
\label{eq:47}
\eeq
These formulas are the continuous analogs of
 (\ref{eq:312}) and (\ref{eq:3l}) respectively.

{\bf Proof of Theorem \ref{thm:51}:}
For a fixed integer $L \geq 2$, choose real numbers
\bd
- \infty = a_0 < a_1 < \ldots < a_{L-1} < a_L = + \infty ,
\ed
\bd
- \infty = b_0 < b_1 < \ldots < b_{L-1} < b_L = + \infty
\ed
such that the semi-open intervals $I_i = (a_{i-1},a_i], J_i = (b_{i-1},b_i]$
satisfy
\bd
\mu(I_i) = 1/L , \nu(J_i) = 1/L , i = 1 , \ldots , L .
\ed
Now define the equiprobable
partition of $\R^2$ consisting of the $L \times L$ grid
$\{ I_i \times J_j , i , j = 1 , \ldots , L \}$.
Next, based on the $l$-length empirical sample $\{ (x_1 , y_1) , \ldots ,
(x_l,y_l) \}$, construct empirical marginal distributions $\hat{\mu}$
for $X$ and $\hat{\nu}$ for $Y$.
Based on these empirical marginals, divide both the $X$-axis $\R$
and $Y$-axis $\R$ into $L$ bins each having nearly equal fractions
of the $l$ samples in each bin.
This gives an empirical $L \times L$ partitioning of $\R^2$,
which is denoted by
$\{ I_i^{(L)} \times J_j^{(L)} , i , j = 1 , \ldots , L \}$.
Using this grid,
compute the associated empirical joint distribution $\hat{\th}_l$ on $\R^2$.
Then the proof of \cite[Lemma 1]{WKV05} can be adapted to show that
the empirical partition
$\{ I_i^{(L)} \times J_j^{(L)} , i , j = 1 , \ldots , L \}$
converges to the true partition
$\{ I_i \times J_j , i , j = 1 , \ldots , L \}$ as $\lai$,
with respect to the product measure $\mu \times \nu$.
The only detail that differs from \cite{WKV05} is the computation of
the so-called `growth function'.
Given a set $A \seq \R^2$ of cardinality $m$, the number of different ways in
which this set can be partitioned by a rectangular grid of dimension
$L \times L$ is called the growth function, denoted by $\D_m$.
It is shown in \cite[Eq. (15)]{WKV05} that when the partition consists
of $L$ intervals and the set being partitioned is $\R$, then
$\D_m$ is given by the combinatorial parameter
\bd
\D_m = \left( \ba{c} m+L \\ L \ea \right) =
\frac{ (m+L)! }{ m! L! } .
\ed
It is also shown in \cite[Eq. (21)]{WKV05} that
\bd
\frac{1}{m} \log \left[ \left( \ba{c} m+L \\ L \ea \right) \right]
\leq 2m h( 1/L) ,
\ed
where $h(\cdot)$ is defined by
\bd
h(x) = - x \log x - (1-x) \log (1-x) , \fa x \in (0,1) .
\ed
When $\R$ is replaced by $\R^2$ and a set of $L$ intervals is
replaced by a grid of $L^2$ rectangles, it is easy to see that the
growth function is {\it no larger than\/}
\bd
\D_m \leq \left[ \left( \ba{c} m+L \\ L \ea \right) \right]^2 .
\ed
Therefore
\bd
\frac{ \log \D_m }{m} \leq 4m h( 1/L) .
\ed
In any case, since $L$, the number of grid elements, approaches
$\infty$ as $\lai$,
it follows that the growth condition proposed in \cite{LN96} is
satisfied.
Therefore the empirical partition converges to the true partition
as $\lai$.

Next, let $\{ I_i \times J_j , i , j = 1 , \ldots , L \}$ denote, as before,
the true equiprobable $L \times L$ gridding of $\R^2$.
Suppose that, after  $l$ samples $(x_r,y_r) , r = 1 , \ldots , l$
have been drawn, the data is put into $k_l$ bins.
Then the expression (\ref{eq:43a}) defining the
true $\beta$-mixing coefficient can be rewritten as
\bd
\beta(X,Y) = 0.5 \sum_{i=1}^{k_l} \sum_{j=1}^{k_l} \int_{I_i} \int_{J_j}
| f(x,y) - 1 | d \mu(x) d \nu(y) .
\ed
Now suppose $l$ is an exact multiple of
$k_l$.
Then the empirical estimate based on the $k_l \times k_l$
empirical grid can be written as
\bd
\betah_l = 0.5 \sum_{i=1}^{k_l} \sum_{j=1}^{k_l} | C_{ij} - 1 | k_l^{-2} ,
\ed
where $C_{ij}$ denotes the number of samples $(x_r,y_r)$ in the $ij$-th cell
of the {\it empirical\/} (not true) equiprobable grid.
If $l$ is not an exact multiple of $k_l$, then some bins will have
$\lfloor l/k_l \rfloor$ elements while other bins will have
$\lfloor l/k_l \rfloor + 1$ elements.
As a result, the term $k_l^{-2}$ gets
replaced by $(s_i t_j)/l^2$ where $s_i$ is the number of samples in 
$I_i^{(l)}$ and $t_j$ is the number of samples in $J_j^{(l)}$.
Now, just as in \cite[Eq.\ (36) {\it et seq.}]{WKV05}, the
error $| \betah_l - \beta (X,Y) |$ can be bounded by the sum of
two errors, the first of which is caused by the fact that the empirical
equiprobable grid is not the same as the true equiprobable grid
(the term $e_1$ of \cite{WKV05}), and the second is the error caused
by approximating an integral by a finite sum over the true equiprobable grid
(the term $e_2$ of \cite{WKV05}).
Out of these, the first error term goes to zero as $\lai$ because,
if $k_l/l \ap 0$ so that each bin contains increasingly
many samples, the empirical equiprobable grid converges to the true
equiprobable grid.
The second error terms goes to zero because the integrand in (\ref{eq:43a})
belongs to $L_1(\R^2,\mu \times \nu)$, as shown in \cite[Eq.\ (37)]{WKV05}.
$\halmos$


{\bf Proof of Theorem \ref{thm:52}:}
The main source of difficulty here is that, whereas the expression for
$\beta(X,Y)$ involves just a single integral, the expressions for
$\al(X,Y)$ and for $\phi(X,Y)$ involve the supremum over all sets $T \seq \R$.
Thus, in order to show that the empirical estimates converge to
the true values, we must show not only that empirical estimates of
integrals of the form
$\int_\R [\kappa_T]_+ d \mu(x)$ and $\int_\R [\d_T]_+ d \mu(x)$
converge to their correct values for each fixed set $T$, but also
that the convergence is in some sense uniform with respect to $T$.
This is where we use the boundedness of the density $f(\cdot,\cdot)$.
The details are fairly routine modifications of arguments in \cite{WKV05}.
Specifically, (switching notation to that of \cite{WKV05}),
suppose that in their Equation (27), we have not just one measure $\mu$,
but rather a family of measures $\mu_T$, indexed by $T$, and suppose
there exists a finite constant $c$ such that for every set $S$ we
have $\mu_T(S) \leq c Q(S)$.
Then it follows from Equation (27) {\it et seq.} of \cite{WKV05} that
\bd
\mu_T ( ( a_i \wedge a_l^m , a_i \vee a_i^m ] )
\leq c Q( ( a_i \wedge a_l^m , a_i \vee a_i^m ] ), \fa T .
\ed
Therefore
\bd
\lim_{\mai} \sup_T \mu_T ( ( a_i \wedge a_l^m , a_i \vee a_i^m ] ) = 0 .
\ed
With this modification, the rest of the proof in \cite{WKV05} can be
mimicked to show the following:
In the interests of brevity, define
\bd
r_T = \int_\R [\kappa_T]_+ d \mu(x) 
\ed
and let $\hat{r}_{T,l}$ denote its empirical approximation.
Then, using the above modification of the argument in \cite{WKV05},
it follows that
\bd
\lim_{\lai} \sup_T | r_T - \hat{r}_T | = 0 .
\ed
As a consequence,
\bd
\lim_{\lai} \sup_T \hat{r}_T = \sup_T r_T = \al(X,Y) .
\ed
The proof for the $\phi$-mixing coefficient is entirely similar.
$\halmos$

\section{Concluding Remarks}\label{sec:7}

In this paper we have studied the problems of computing and estimating
the mixing coefficients between two random variables in two important
cases, namely: finite-valued and real-valued random variables.
Three different mixing coefficients were studied, namely $\al$-mixing,
$\beta$-mixing and $\phi$-mixing coefficients.
In the case of finite-valued random variables,
it has been shown that determining whether the $\al$-mixing coefficient 
exceeds a prespecified threshold is an NP-complete problem.
Efficiently computable upper and lower bounds for the $\al$-mixing
coefficients have been derived.
In contrast, an explicit and efficiently computable formula has been
derived for the $\phi$-mixing coefficient.
Analogs of the data-processing inequality from information theory have
been established for each of the three kinds of mixing coefficients.
In the case of real-valued random variables,
by using percentile binning and allowing the number of bins
to increase more slowly than the number of samples, we can generate
empirical estimates that converge to the true values for all the three
kinds of mixing coefficients.

Several interesting questions are thrown up by the contents of this paper.
As mentioned in the introduction, mixing coefficients were originally
introduced as a way of extending the law of large numbers to stochastic
processes that are not i.i.d.
The problem studied in Section \ref{sec:5} is to estimate the mixing
coefficient between two real-valued random variables $X$ and $Y$, based
on i.i.d.\ samples of the pair.
A counterpoint to this problem is that studied in \cite{MSS11}, where
the objective is to estimate the $\beta$-mixing rate of
a stationary stochastic process $\{ X_t \}$ from a single sample path.
For a fixed integer $k$, the rate $\beta(k)$ can be interpreted as
the $\beta$-mixing coefficient between the semi-infinite past $X_{-\infty}^0$
and the semi-infinite future $X_k^\infty$.
However, the techniques presented here do not work for that problem,
whereas \cite{MSS11} presents a comprehensive solution in terms of
``blocking'' and ``histogramming'', that is, estimating the joint
distribution of $d_l$ consecutive random variables, when $l$ samples
are available in all.
It is interesting to note that the convergence results in \cite{MSS11}
also depend on letting $d_l$ grow more slowly than $l$.
Specifically, as shown in \cite[Theorem 2.3]{MSS11}, the estimator
converges to the true value provided $d_l = O(\exp[W(\log l)])$,
where $W$ is the Lambert $W$ function.
More details can be found in \cite{MSS11}.
It would be worthwhile to explore whether similar estimators can be
constructed for the $\al$-mixing rate of a stochastic process.

Another direction is to explore whether analogs of the data processing
inequality, namely (\ref{eq:64}) through (\ref{eq:67}), hold for real-valued
random variables.
The proof techniques in Section \ref{sec:6} make heavy use of the finiteness
of the underlying sets where the various random variables assume their values.
On the other hand, there are analogous formulas for real-valued random
variables, namely (\ref{eq:46}) and (\ref{eq:47}).
It might therefore be possible to extend the proofs in Section \ref{sec:6}
making use of these formulas.
However, the technicalities may prove to be formidable.

In the consistency theorems of Section \ref{sec:5}, the requirement
that the bins should consist of empirically equiprobable (or percentile)
samples is really not necessary.
A close examination of the proof techniques used in \cite{WKV05} shows
that, so long as the minimum number of samples in each bin approaches
infinity as $\lai$, the results would still hold.
We leave it to the reader to state and prove such results.
The later parts of the paper \cite{WKV05} contain some proposals on
how to speed up the convergence of the empirical estimates
of the Kullback-Leibler divergence between two unknown measures.
It would be worthwhile to explore whether similar speed-ups can
be found for the algorithms proposed here for estimating mixing coefficients
from empirical data.

\end{document}